# How to spot outliers: an Ensemble Anomaly Detection Framework

**Daniil Peysakhovich**
*UBS, QFAI Institute*
daniil.peysakhovich@ubs.com

**Rafał Sieradzki**
*NYU Stern, QFAI Institute, Krakow University of Economics*
rjs9362@nyu.edu
*Corresponding author*

---

## Abstract

Errors in risk valuation outputs – arising from data-feed failures, model misconfiguration, or system malfunctions – can propagate undetected through an investment bank's risk infrastructure and generate material operational losses. Using proprietary daily credit-derivatives data from a major global investment bank covering 183 trades across 129 trading days, we design, implement, and empirically evaluate the Ensemble Quality Assessment Framework (EQAF): a layered, unsupervised architecture that combines complementary outlier-detection methods to monitor risk calculation integrity in real time. Using a controlled anomaly injection protocol with eight operationally realistic scenarios, we demonstrate that the calibrated ensemble achieves F1-scores of 61–79% – substantially outperforming the best individual method (6–66%) across four distinct risk-measure datasets. AUC-ROC improvements of 4–6 percentage points confirm the advantage is robust to threshold selection. We further establish that purely statistical detection methods structurally fail to identify stale-value anomalies – a class of frozen-feed errors in which valuation outputs are numerically identical to prior observations and therefore indistinguishable from normal data – and that domain-specific deterministic rules are architecturally indispensable. Our findings have direct implications for model risk management under Basel III and the Fundamental Review of the Trading Book (FRTB), where automated, auditable quality controls over internal risk models are increasingly mandated.

---

## 1. Introduction

The integrity of risk calculation systems sits at the heart of sound banking operations. Modern investment banks rely on complex valuation engines to produce daily risk metrics – Value-at-Risk (VaR), Expected Shortfall (ES), credit sensitivities, and stress losses – that underpin capital allocation, hedging decisions, regulatory reporting, and counterparty risk management. When these engines malfunction, the consequences can be severe and, critically, can remain hidden for extended periods. The 2012 "*London Whale*" incident at JPMorgan Chase – in which a misconfigured VaR model understated the risk of a multi-billion-dollar credit-derivatives portfolio for months before losses exceeded $6 billion – is perhaps the most prominent illustration (Zeissler et al., 2019). The 2008 Société Générale rogue-trading scandal, where $4.9 billion in losses accumulated partly because risk monitoring systems failed to flag abnormal valuation patterns, and the 2012 Knight Capital algorithmic trading failure, which generated $440 million in losses within 45 minutes due to system misbehavior, reinforce the same message: errors within risk-calculation and trading systems are material sources of operational risk.

These events are no longer treated as idiosyncratic bad luck. Berger et al. (2022) demonstrate empirically, using supervisory data from large U.S. bank holding companies (BHCs), that operational losses elevate systemic risk through two channels: a direct channel that impairs market values of affected institutions, and a correlated-loss channel through which multiple institutions are hit simultaneously. Chernobai et al. (2021) document that business complexity – the defining feature of derivatives desks – amplifies operational-risk exposure. Curti et al. (2022) find that the largest BHCs bear disproportionately high operational losses per unit of assets. Taken together, this literature establishes that operational failures within trading and risk-measurement operations are not merely internal management problems: they threaten the stability of the broader financial system.

At the same time, global banking regulators have responded to this evidence by strengthening requirements for model oversight. The Basel Committee on Banking Supervision's SR 11-7 framework (Board of Governors of the Federal Reserve System, 2011) mandates comprehensive governance, documentation, and ongoing monitoring of all risk models. The Fundamental Review of the Trading Book (FRTB) requires banks using the Internal Models Approach (IMA) to continuously monitor the performance of their Expected Shortfall models through mechanisms such as backtesting and Profit and Loss Attribution (PLA) testing. Persistent failures may result in the loss of IMA eligibility at the trading-desk level, creating strong incentives for ongoing model-performance monitoring and quality-control processes. (Basel Committee on Banking Supervision, 2019). The European Banking Authority's 2025 update to operational-risk reporting standards further demands that valuation-process failures be systematically identified, categorized and reported (European Banking Authority, 2025). These regulatory pressures create direct institutional incentives – and legal obligations – for automated, auditable quality assessment of risk calculation outputs.

Yet, despite the regulatory demand and the growing literature on operational risk, the specific problem of systematic, automated quality control for risk valuation outputs remains relatively underexplored in academic literature. Existing work on financial anomaly detection focuses predominantly on fraud detection in transactions (Hilal et al., 2022; Bakumenko and Elragal, 2022), payment-system monitoring (Desai et al., 2024), and credit-risk modelling (Leo et al., 2019). The anomaly-detection methods literature (Chandola et al., 2009; Aggarwal, 2017; Liu et al., 2008) is rich in algorithmic development but does not address governance, interpretability, and operational constraints specific to investment-banking risk systems. Schmidt et al. (2023) and Ding et al. (2023) propose an automated quality-control system for traceable and reproducible environmental data streams, while related work on IoT data-quality pipelines develops domain-specific mechanisms for assessing streaming sensor data. These frameworks demonstrate the relevance of automated data-quality control, but they are not designed for derivatives risk outputs or for the regulatory and statistical constraints associated with market-risk model monitoring..

This paper aims to address that gap. We develop and empirically evaluate the Ensemble Quality Assessment Framework (EQAF) – a modular, unsupervised architecture that monitors the integrity of investment-banking risk calculations on a trade-by-trade, day-by-day basis. EQAF allows integrating complementary statistical and machine-learning detection methods – for experiments we used Robust Z-Score, Empirical CDF scoring, Hampel Filter, Rolling Z-Score, Local Outlier Factor (LOF), and Isolation Forest (IF) – with a deterministic stale-value filter, calibrated score normalization using empirical quantile ranking, and a weighted aggregation layer with qualified-majority voting. The layered design follows the theoretical insight of Aggarwal (2013, 2015) that independent ensembles outperform individual methods by reducing variance and broadening anomaly coverage.

We evaluate EQAF using proprietary data from a major global investment bank (UBS Investment Bank), covering 183 credit-derivative trades – including Single Name CDS, Index CDS, Bond Basis Swaps, Basket CDS, and Index Tranche CDS – across six months of daily observations. Four datasets were analyzed: 1 – Present Value, 2 – P&L slices, 3 – Credit Delta on Single Curve shift, 4 – Credit Delta on Index Curve shift. To overcome the absence of labelled production anomalies, we deploy a controlled injection protocol with eight operationally realistic scenarios: PointShock, ScaleError, SignFlip, StaleValue, Drift, ClusterShock, TradeTypeWideShock, and SuddenZero[1]. These scenarios are calibrated to the median absolute deviation (MAD) of each trade-feature series and injected exclusively into the out-of-sample period, ensuring clean separation between training and evaluation.

Our principal findings are fourfold. First, the calibrated ensemble achieves F1-scores of 61–79% across four risk-measure datasets, consistently and substantially outperforming the best individual method (6–66%). AUC-ROC improvements of 4–6 percentage points over the best standalone method confirm the advantage is robust to threshold selection. These gains are synergistic, not merely additive: certain anomaly types captured by the ensemble are entirely missed by each individual method in isolation. Second, we establish a structural result: all six statistical detection methods completely fail to detect StaleValue anomalies (0% recall), because a frozen value is by construction statistically indistinguishable from normal observations. The deterministic filter raises this to 100%, demonstrating that some quality failure modes require domain-specific rule encoding rather than distributional learning. Third, detection difficulty varies systematically across datasets in a manner consistent with their statistical properties: Credit Delta Index, the most homogeneous dataset, yields the highest ensemble AUC (0.979), while the high-volatility P&L Slices dataset is the most challenging (AUC 0.836), providing a risk-sensitivity gradient for practitioners. Fourth, the results indicate that a framework-based approach is necessary: model calibration and tuning must be dataset-specific, as no single configuration can be expected to perform reliably across all datasets.

[1] These scenarios will be described in section 3.3.

These results carry direct implications for model risk management. They demonstrate that automated, scalable quality control of risk-calculation outputs is technically feasible using available detection methods, but requires careful ensemble composition, domain-aware rule augmentation, and dataset-specific calibration. The framework is consistent with the explainability and auditability requirements of SR 11-7 and FRTB, providing interpretable scoring signals that can be integrated into existing model risk governance workflows.

The remainder of the paper proceeds as follows. Section 2 reviews the relevant literature on operational risk, anomaly detection, and data quality management. Section 3 describes the institutional context and dataset. Section 4 develops the EQAF methodology in detail. Section 5 presents empirical results. Section 6 discusses implications for practice and regulation. Section 7 concludes. An Appendix provides additional results for all four datasets.

## 2. Institutional background and related literature

### *2.1 Operational Risk in Investment Banking*

The modern study of operational risk in banking has moved from a primarily regulatory exercise to a substantive empirical literature. Following the original Basel II definition – "the risk of loss resulting from inadequate or failed internal processes, people and systems or from external events" (Basel Committee on Banking Supervision, 2006) – early academic work established that operational losses at financial institutions were larger, more frequent and more capital-consuming than generally appreciated. De Fontnouvelle et al. (2006) provide early evidence using internal loss databases and show that extreme operational losses are heavy-tailed and that capital requirements derived from standard risk models can substantially understate true exposure.

More recently, literature has moved toward understanding the determinants and systemic consequences of operational risk. Berger et al. (2022) is the most direct antecedent to our work: using supervisory data from large U.S. BHCs, they show that operational losses increase systemic risk – measured by SRISK, ΔCoVaR and SES – through both a direct market-value impairment channel and a correlated-loss channel. Their finding that tail events drive these effects is particularly relevant to quality failures in risk-calculation systems, which typically manifest as rare but large discrepancies. Chernobai et al. (2021) show that business complexity – exactly the characteristic of large derivatives operations – elevates operational risk exposure. Curti et al.

(2022) document size effects, finding that the largest BHCs bear higher operational losses per dollar of assets, largely driven by client-service and execution failures.

Within the taxonomy of Basel Committee on Banking Supervision (2003), our paper focuses specifically on the "execution, delivery and process management" event type, applied to the trading and sales business line. This encompasses data-entry errors, valuation failures, incorrect model implementation, and system-level deficiencies that affect the integrity of risk metrics. Hull (2018) and Jorion (2007) both emphasize that reliable risk measurement is not a sufficient condition for sound risk management – it is a precondition. Inaccurate VaR or ES figures mislead management, enable excessive risk-taking, and distort regulatory capital calculations. Jorion (2007) specifically notes that VaR models must be validated through back-testing precisely because output errors, if undetected, create false confidence. Our paper operationalizes this validation requirement as a continuous, automated monitoring process.

The 2012 JPMorgan London Whale case provides a canonical illustration of how execution failures in risk calculation systems materialize into major losses. Zeissler et al. (2019) document that the VaR model used to monitor the Chief Investment Office's credit-derivatives portfolio silently producing numbers approximately half the true risk exposure for several months. This failure was not discovered through manual review or existing controls – it required an independent external assessment. The case motivated both academic interest in model quality control and regulatory responses including SR 11-7 enhancements and FRTB desk-level IMA approval requirements.

### *2.2 Anomaly Detection: Methods and Applications in Finance*

The anomaly detection literature provides the technical foundation for our framework. Chandola et al. (2009) survey the field comprehensively, distinguishing point anomalies (individual deviations), contextual anomalies (observations that are unusual given their temporal or spatial context), and collective anomalies (groups of observations that are anomalous together). All three types arise in risk-calculation outputs: a miscalibrated scaling factor produces point anomalies; a stale-value feed produces contextual anomalies (the value is normal in magnitude but abnormal given market movements); a correlated system failure produces collective anomalies across multiple trades simultaneously. Chandola et al. (2009) conclude that no single technique is universally optimal and that performance is highly dependent on data characteristics – a finding that motivates our ensemble approach.

Aggarwal (2017) provides a comprehensive taxonomy of outlier analysis methods, covering distance-based, density-based, clustering-based, statistical, and machine-learning approaches. Critical for our design, Aggarwal (2013, 2015) formalizes the concept of *outlier ensembles*: aggregations of multiple detection algorithms whose combined output substantially outperforms any constituent method. Aggarwal (2015) demonstrates theoretically that independent ensembles – where methods operate in parallel on the same data and outputs are aggregated – reduce variance in anomaly scores and improve coverage across heterogeneous anomaly types, even when each individual method has blind spots. In addition to independent (parallel) ensembles, Aggarwal introduces sequential ensembles. This theoretical framework directly motivates EQAF design which combines both independent and sequential flows.

The Isolation Forest algorithm (Liu et al., 2008) is now a widely adopted unsupervised anomaly detection method. By recursively partitioning the feature space using random cuts, Isolation Forest assigns higher anomaly scores to observations that require fewer cuts to isolate – an efficient proxy for low-density outliers. Local Outlier Factor (LOF), introduced by Breunig et al. (2000), quantifies the density-based outlierness of each observation relative to its neighbors, making it effective for detecting contextual anomalies in heterogeneous datasets. Both methods are included in EQAF for their complementary coverage: Isolation Forest handles global density anomalies efficiently in high-dimensional settings, while LOF captures local context-dependent deviations.

Within financial literature, anomaly detection has been applied primarily to fraud, payments, and accounting. Hilal et al. (2022) survey anomaly detection techniques for financial fraud, covering statistical, distance-based, clustering, and ensemble methods, and document that hybrid ensemble approaches systematically outperform single-method baselines in imbalanced real-world settings – consistent with our findings. Bakumenko and Elragal (2022) demonstrate the practical feasibility of unsupervised anomaly detection in financial accounting data, showing that both Isolation Forest and autoencoders can identify abnormal journal entries with recall rates exceeding 90% on carefully prepared datasets. Crépey et al. (2022) propose a sequential PCA–neural-network ensemble for anomaly detection in financial time series, demonstrating that dimensionality reduction followed by reconstruction-error detection improves robustness in correlated, high-dimensional settings.

Recent work has extended anomaly detection to payment-system monitoring. Desai et al. (2024), in an important paper closely related to ours, apply machine-learning methods to detect anomalies in high-value payment systems in real time. They find that unsupervised ensemble approaches substantially outperform rule-based baselines in detecting temporally structured anomalies, and that detection performance is highly sensitive to the temporal ordering of data – a finding that motivates our inclusion of rolling-window and stateful methods alongside static distributional measures.

### *2.3 Data Quality Management Frameworks*

The data quality literature provides the conceptual framework for how EQAF outputs should be interpreted and integrated into governance structures. Pipino et al. (2002) establish the foundational framework for data quality assessment, distinguishing objective measurement (derived from the data itself), subjective assessment (based on user experience), and context-dependent criteria (fitness-for-purpose). They introduce the concept of data quality dimensions – including accuracy, completeness, consistency, and timeliness – which map directly to the anomaly types we consider: inaccurate valuations (accuracy), missing market data (completeness), inconsistent risk-factor mappings (consistency), and stale feeds (timeliness). Batini and Scannapieco (2016) extend this framework to large-scale data management, introducing structured methodologies for quality assessment that emphasize repeatable, metadata-driven rule systems.

Schmidt et al. (2023) propose SaQC, a modular pipeline for automated quality control of time-series data streams in environmental science. SaQC formalizes quality assessment as a configurable sequence of checks that generate flags rather than modifying underlying data – an architectural principle we adopt directly. Their emphasis on traceability, reproducibility, and the separation of detection logic from data storage aligns closely with the governance requirements of investment-banking risk systems. Ding et al. (2023) extend the quality-management perspective to a human-in-the-loop framework that explicitly separates machine-generated anomaly explanations from human-driven reasoning and remediation. Their insight that automated detection signals are necessary but not sufficient for effective quality management – and that interpretable, actionable outputs are essential – informs our scoring and flagging architecture.

### *2.4 Regulatory Context and Model Risk Management*

The regulatory environment for model risk in investment banking has evolved substantially since the financial crisis. The Federal Reserve's SR 11-7 guidance (Board of Governors of the Federal Reserve System, 2011) established the requirement for comprehensive model risk management, including independent validation, ongoing monitoring, and documentation of all models used for risk measurement. More recently, the Fundamental Review of the Trading Book framework (FRTB , Basel Committee on Banking Supervision, 2019) has raised the bar for internal risk models by requiring desk-level IMA approval, Expected Shortfall calibration to stressed periods, Non-Modellable Risk Factor treatment, and ongoing Profit and Loss Attribution testing. Institutions whose models fail these tests are required to revert to the standardized approach, which carries substantially higher capital requirements.

European Banking Authority (EBA, 2025) has recently strengthened operational-risk event reporting requirements, including a new taxonomy for valuation-process failures. These regulatory developments create a clear external mandate for the type of automated quality assessment that EQAF provides. The Financial Stability Board (2014) has additionally emphasized that risk culture at systemically important banks must be assessed regularly, specifically including the reliability of internal information systems. Hull (2018) synthesizes these requirements into a practitioner perspective, arguing that effective risk culture requires systematic quality controls at every stage of the risk measurement and reporting process.

Our paper contributes to this regulatory and academic context by providing the first rigorous empirical demonstration that ensemble anomaly detection can serve as an automated, scalable quality-control mechanism for investment-banking risk calculations. We bridge the gap between the anomaly-detection methods literature, the operational risk management literature, and the regulatory model-risk management framework – a gap that, to our knowledge, has not been previously addressed.

## 3. Data and Institutional Context

### *3.1 Risk Valuation Architecture in Investment Banking*

A risk valuation system is the analytical core of an investment bank's trading operation. It combines three principal inputs – trade data (notional, trade type, maturity, counterparty), market data (yield curves, credit spreads, volatility surfaces, exchange rates), and model configurations (pricing models, calibration parameters, risk-factor mappings) to produce daily risk metrics for each instrument in the trading book. These metrics serve two parallel purposes: internal risk management (limit monitoring, hedging, capital allocation) and regulatory reporting (VaR and ES under Basel III, sensitivities and capital charges under FRTB).

Figure 1 illustrates the principal architecture of a risk valuation system and the error propagation pathway that motivates our research. Outputs from the valuation engine flow to multiple users simultaneously – Market Risk Control, Finance, Regulatory Capital, Counterparty Credit Risk, Middle Office, and Internal Audit – often with no standard mechanism for cross-validating output integrity before distribution. This architecture creates a material propagation risk: a single error in the valuation engine, if undetected, affects all downstream users simultaneously.

***Figure 1. Principal risk valuation system architecture and error propagation pathway in an investment bank.***

The risk metrics that are most relevant for this study are credit sensitivities (Delta) and valuations (PV, P&L). Credit Delta measures the change in present value of a credit instrument for a one basis-point shift in the underlying credit spread – a first-order sensitivity used to hedge and monitor credit risk exposures. Present Value (PV) reflects the current mark-to-market value of each trade. Profit and Loss (P&L) attribution decomposes daily valuation changes into economically meaningful components driven by different market risk factors. These and other metrics provide a comprehensive view of a trading portfolio's credit risk profile and are among the most operationally critical outputs of the valuation system.

Under the FRTB Internal Models Approach, banks are required to maintain ongoing evidence that internal market risk models remain suitable for regulatory capital purposes. In particular, the framework requires trading desk-level Profit and Loss Attribution tests, which assess the consistency between risk-theoretical P&L and hypothetical P&L, as well as back-testing, which evaluates the accuracy of model-based VaR forecasts against realized outcomes. While these tests are distinct from automated output-level quality control, they reflect a broader regulatory expectation that risk model outputs should be subject to systematic and continuing validation. In this context, EQAF can be positioned as a complementary quality-assessment layer, supporting earlier detection of anomalous valuation or risk outputs before they affect downstream model monitoring, reporting, or capital processes. As the Basel Committee on Banking Supervision (2019) specifies, failure to satisfy desk-level eligibility tests may lead to loss of Internal Models Approach (IMA) approval for a trading desk, thereby creating a direct capital incentive for proactive monitoring.

### *3.2 Dataset Description*

Our empirical analysis employs real risk-calculation output from UBS Investment Bank, provided under a confidentiality agreement for research purposes. The dataset covers 183 credit-derivative trades across six months of daily calculations from 1 January to 30 June 2025, yielding 129 trading days of observations. Confidential trade details were removed or obfuscated prior to extraction; data samples are drawn from a period already disclosed in the bank's public financial statements, ensuring no proprietary information is revealed.

The dataset (Table 1) covers five credit-derivative trade types: (1) Single Name Credit Default Swap (CDS) – bilateral contracts providing credit protection against a single reference entity; (2) Index CDS – standardized protection on a basket of reference entities (iTraxx in Europe, CDX in the U.S.); (3) Bond Basis Portfolio Swap – structured derivatives exploiting or hedging differences between the bond market and CDS market for the same issuer; (4) Basket CDS – protection on a predefined basket of reference entities with contingent payoffs; and (5) Index Tranche CDS – structured credit derivatives providing exposure to a specific loss segment of a CDS index portfolio. This trade-type diversity creates the heterogeneous statistical properties of the dataset and mirrors the actual composition of a sample investment bank's credit derivatives book.

***Table 1. Dataset Summary Statistics***

| Dataset | Trades | Features | Obs. (train) | Obs. (OOS) | Volatility |
|---|---|---|---|---|---|
| Credit Delta Single | 183 | 1 | 15,870 | 5,486 | Low |
| Credit Delta Index | 52 | 1 | 8,944 | 3,126 | Low–Medium |
| Present Value (PV) | 183 | 2 | 16,155 | 5,465 | High |
| P&L Slices | 183 | 7 | 16,287 | 5,465 | Very high |

*Notes: "Trades" is the count of unique trade IDs in each dataset; Credit Delta Index covers only trades with index curve exposure. "Features" is the number of numerical columns per trade-date observation. OOS = out-of-sample period (trading days 90–129, from 16 May 2025).*

The four datasets, illustrated on Figure 2, exhibit substantially different statistical properties, which creates a natural gradient of detection difficulty for our empirical evaluation. Credit Delta datasets show near-linear dynamics with moderate day-to-day variation characteristic of mark-to-model sensitivities. The PV dataset introduces higher natural volatility, including a pronounced spike during the US tariff announcement in April 2025 (which appears in the training period and provides a realistic test of whether the framework generates excessive false positives during legitimate market stress). The P&L dataset is the most challenging, with high baseline volatility, non-Gaussian distributions, and frequent large legitimate movements driven by market dynamics.

***Figure 2. Stylized illustration of daily time series for all four risk-measure datasets***

Figure 2. Overview of the Four Risk-Measure Datasets with Injected Anomalies (OOS Period)

*Notes: illustration covers date between 1 January – 30 June 2025. Training period (Days 1–89); OOS period (Days 90–129) with injected anomalies shown in red. Dashed line marks the training/OOS boundary. Scales reflect relative magnitudes within each dataset.*

### *3.3 Anomaly Injection Protocol*

Because the production dataset contains no verified quality failures, evaluating anomaly detection performance requires a controlled injection protocol. Following the approach of Aggarwal (2017) and Chandola et al. (2009), we inject artificial anomalies exclusively into the OOS period, leaving the training period clean for parameter estimation. This design ensures that all calibration is performed under realistic conditions (no leakage of anomaly labels) and that evaluation reflects genuine out-of-sample performance.

Table 2 describes the eight injection scenarios, which were designed to cover the principal failure modes observed in production derivatives risk systems based on operational risk event records and domain knowledge. These scenarios are also illustrated on Figure 3. Each scenario is calibrated using the median absolute deviation (MAD) of the training-period values for each trade-feature series, with a severity coefficient k[2] to ensure injections are operationally meaningful. Labels are attached to injected records in the raw data but are not passed to EQAF at any stage, preventing any information leakage during training, scoring, or calibration.

***Table 2. Anomaly Injection Scenarios***

| Scenario | Scope | Selection | Formula | Explanation |
|---|---|---|---|---|
| **PointShock** | All datasets | Random records | $v' = v + k \cdot tMAD$ | One-day bad snapshot / transient valuation glitch. Missing market data on one curve Failed calibration run Partial revaluation |
| **ScaleError** | PV, Credit Deltas | Random records | $v' = \begin{cases} v \cdot scale, & 50\% \\ \frac{v}{scale}, & 50\% \end{cases}$ | Unit, currency or notional scaling error, decimal placement bug |
| **SignFlip** | All datasets | Random records | $v' = -v$ | Sign convention inversion, wrong direction, mapping inversion bug |
| **StaleValue** | All datasets | Random trades, 3 days in a row | $v^{t\prime} = v^{t-1}$ | Valuations not updating, feed stuck, source frozen |
| **SuddenZero** | PV, Credit Deltas | Random records | $v' = 0$ | Valuation error, dead code path, null/missing value, convention inversion |
| **Drift** | All datasets | Random trades, 10+ consecutive days | $v'_i = v_i + rs \cdot \frac{i}{T-1} k \cdot tMAD,$ $i = 0, 1, \cdots, T-1$ | Gradual model degradation, stale curve, wrong rates assumption |
| **TradeType WideShock** | All datasets | Subset of trades, 1 day | $v' = v + rs \cdot k \cdot tMAD$ | Market-wide repricing event affecting risk category, credit curve shift |
| **CluserShock** | All datasets | Random trades, 3 days in a row | $v' = v + rs \cdot k \cdot tMAD$ | Market event/shock spanning multiple days, rate spike, credit events |

*Notes: k = 6 (severity multiplier); tMAD = median absolute deviation of training values for each trade-feature series; rs = random sign (±1 with equal probability); scale = $10^2$ for trades with |v| > 1, $10^3$ for |v| < 1. ScaleError and SuddenZero were not applied to P&L Slices dataset, as these injections are indistinguishable from natural feature behavior in high-volatility P&L series.*

[2] Severity coefficient was set to "6.0" within the tests. This value was determined empirically to maintain required level of detectability

*Figure 3. 8 Outlier injection scenarios.*

*Note: Values on the picture are abstract.*

## 4. The Ensemble Quality Assessment Framework (EQAF)

### *4.1 Design Principles*

EQAF is built around four architectural principles, each grounded in literature. First, modularity: each detection method is encapsulated as an independent layer with a standardized scoring interface, enabling method substitution, sensitivity adjustment, and per-dataset calibration without structural changes to the framework. Second, measure-specific segmentation: because Credit Delta, PV and P&L measures differ in economic interpretation and statistical properties, each is processed within a homogeneous subset (Batini and Scannapieco, 2016). Mixing heterogeneous measures into a single analysis would introduce structural missingness and distort detection signals. Third, ECDF-based score normalization: to enable meaningful aggregation across methods with different raw score scales and distributions, each method's outputs are mapped to empirical quantile ranks using the training set. Fourth, hybrid ensemble aggregation: normalized scores are combined using weighted averaging with a qualified-majority override, empirically calibrated against labelled OOS data.

Figure 4 illustrates the full EQAF processing architecture. The flow proceeds in seven steps: (1) data ingestion and feature normalization using training-set denominators; (2) train/OOS split; (3) for each OOS date, apply the deterministic stale-value filter as a sequential pre-screen; (4) pass non-stale observations through all six statistical methods in parallel (described in details in section 4.2); (5) apply ECDF-based score normalization; (6) compute weighted aggregated score with majority-override; (7) generate Green/Amber/Red quality flags and output.

***Figure 4. EQAF processing architecture***

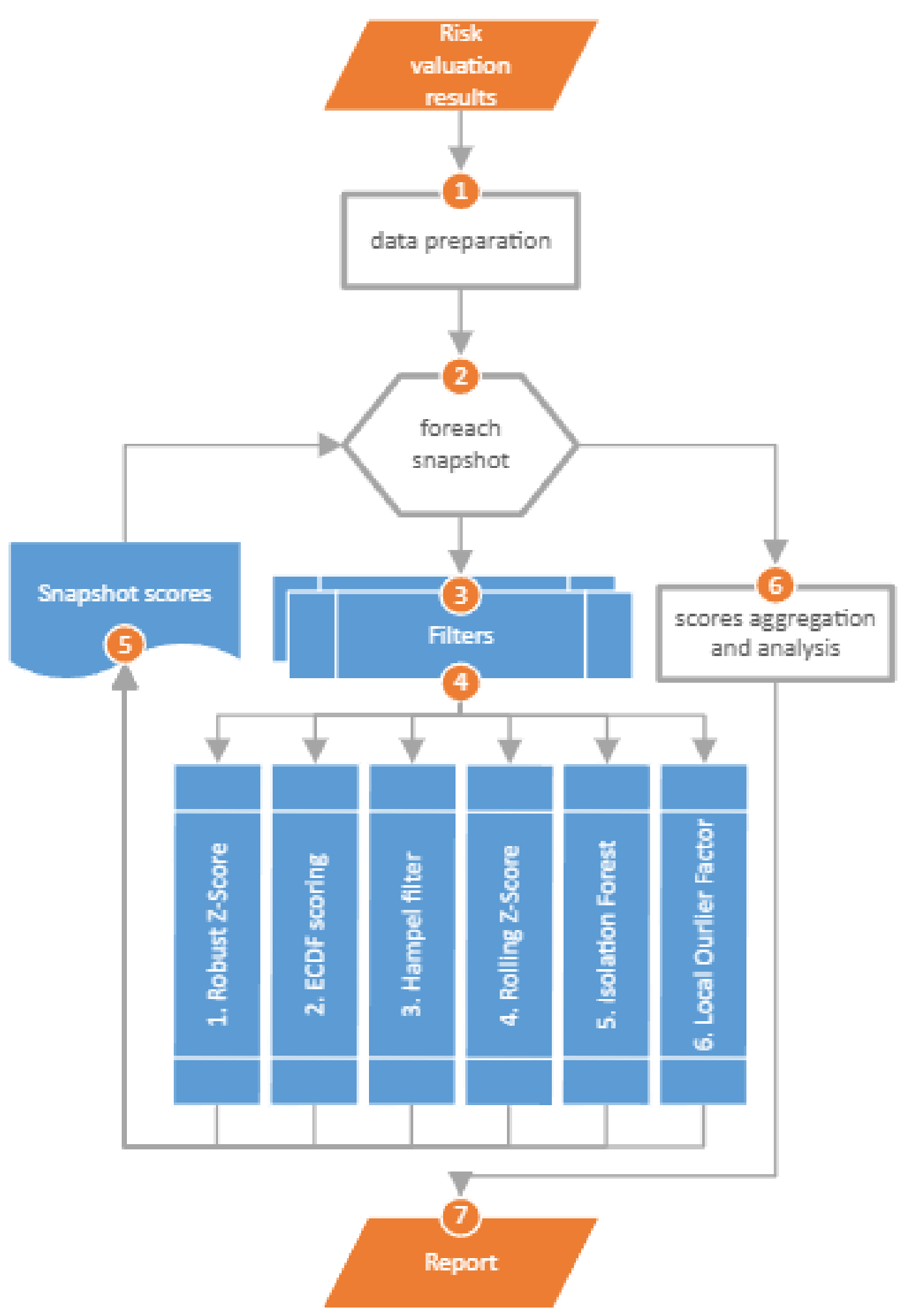

*Notes: The deterministic stale-value filter (Step 3) operates as a sequential pre-screen and overrides the ensemble if triggered. Steps 4–5 implement the independent ensemble: parallel execution, normalization, and aggregation. Set of independent models shown are sample: different models to be used for different implementations of the framework*

## *4.2 Detection Methods*

From the broad set of available outlier detection techniques, this study selects six statistical methods designed to provide complementary detection capabilities. The selection spans four dimensions: method family, statefulness, dimensionality, and temporal dependency. Specifically, the methods cover statistical, distributional, temporal, and density-based approaches; stateless, windowed, and fully stateful designs; univariate and multivariate settings; and both independent and temporally dependent observations. Table 3 summarizes the key properties of each method.

**Table 3. Detection Methods in the EQAF Ensemble**

| Method | Family | Stateful | Dim. | Temporal | Min. Train | Anomaly Types Targeted |
|---|---|---|---|---|---|---|
| Robust Z-Score | Statistical | No | Uni. | None | 20–30 | Global deviations from training distribution; heavy-tailed observations |
| ECDF Scoring | Distributional | Yes | Uni. | Window | 30–50 | Extreme quantile observations; distributional regime shifts |
| Hampel Filter | Statistical | Window | Uni. | Window | 30–50 | Local temporal deviations; isolated spikes within rolling window |
| Rolling Z-Score | Statistical | Window | Uni. | Window | 30–50 | Deviations from recent local mean; temporary shifts; short-term anomalies |
| Isolation Forest | ML-based | Yes | Multi. | None | 50–100 | Global isolation anomalies; multivariate non-linear patterns |
| Local Outlier Factor (LOF)[3] | Density-based | Yes | Multi. | None | 30–50 | Local density anomalies; observations in sparse neighborhoods |

*Notes: "Dim." = dimensionality (Uni. = univariate trade-by-trade; Multi. = multivariate across features or time window). "Temporal" refers to the method's sensitivity to time ordering. "Min. Train" is the approximate minimum training sample size in observations per trade. All methods are unsupervised and trained exclusively on the training period data.*

### 4.2.1 Robust Z-Score

The Robust Z-Score computes the deviation of each observation from the training-period median, scaled by the Median Absolute Deviation (MAD):

$$RZ_t = \frac{x_t - median(X_{train})}{1.4826 \times MAD(X_{train})} \qquad (1)$$

[3] Local Outlier Factor method was implemented but then excluded from all the ensembles during calibration as not efficient for processing data sets.

where $x_t$ is the observed value at time $t$, and $X_{\text{train}}$ denotes the training-period observations. The scaling factor 1.4826 normalizes the MAD to be a consistent estimator of the standard deviation under a normal distribution. This stateless, distribution-free method provides effective detection of global deviations but has no sensitivity to local context or temporal ordering.

### 4.2.2 Empirical CDF Scoring

The Empirical cumulative distribution function (ECDF) method constructs a non-parametric empirical cumulative distribution function from the training data for each trade-feature series. For an observation $x_t$, the empirical CDF is defined as:

$$\widehat{F}(x_t) = \frac{1}{n}\sum_{i=1}^{n} 1(x_i \le x_t) \qquad (2)$$

The anomaly score is then computed as a two-tailed tail probability:

$$S_{\text{ECDF}}(x_t) = min\left(\widehat{F}(x_t), 1 - \widehat{F}(x_t)\right) \qquad (3)$$

where $S_{\text{ECDF}}(x_t) \in [0{,}0.5]$. Lower values indicate observations located further in the tails of the empirical training distribution. To account for distributional drift, the ECDF is updated using a sliding window that incorporates newly observed out-of-sample values.

### 4.2.3 Hampel Filter

The Hampel Filter applies a sliding-window robust outlier test using the rolling median and Median Absolute Deviation (MAD). For a window of width $w$ centered on day $t$, the local median is defined as:

$$\tilde{x}_t = median\left(x_{t-w}, x_{t-w+1}, \ldots, x_t, \ldots, x_{t+w-1}, x_{t+w}\right) \qquad (4)$$

The corresponding local MAD is:

$$MAD_t = median\left(/\, x_i - \tilde{x}_t \,/ : i = t-w, \ldots, t+w\right) \qquad (5)$$

The Hampel statistic is then computed as:

$$H_t = \frac{/\, x_t - \tilde{x}_t \,/}{1.4826 \cdot MAD_t} \qquad (6)$$

An observation is flagged as anomalous when: $H_t > k_H$. Equivalently, the flagging rule can be written as:

$$|x_t - \tilde{x}_t| > k_H \cdot 1.4826 \cdot MAD_t \quad (7)$$

where $k_H$ is a sensitivity threshold calibrated empirically. The Hampel Filter is particularly effective for detecting observations that are inconsistent with their immediate temporal neighbourhood, capturing short-duration anomalies that may not deviate significantly from the global training distribution.

### 4.2.4 Rolling Z-Score

The Rolling Z-score extends the standard Z-score to a moving window by computing the standardized deviation of each observation from its recent rolling mean. For a trailing window of width $w$, the rolling mean is defined as:

$$\mu_t^{\text{roll}} = \frac{1}{w} \sum_{i=t-w}^{t-1} x_i \quad (8)$$

and the rolling standard deviation is:

$$\sigma_t^{\text{roll}} = \sqrt{\frac{1}{w-1} \sum_{i=t-w}^{t-1} \left(x_i - \mu_t^{\text{roll}}\right)^2} \quad (9)$$

The Rolling Z-score is then computed as:

$$RZS_t = \frac{x_t - \mu_t^{\text{roll}}}{\sigma_t^{\text{roll}}} \quad (10)$$

where $\mu_t^{\text{roll}}$ and $\sigma_t^{\text{roll}}$ are estimated over the trailing window preceding observation $x_t$. Unlike the global Robust Z-score, due to rolling context (window) this method detects observations that are anomalous relative to recent behaviour, making it effective for identifying structural-shift anomalies – observations that fall within the global historical range but deviate from the local temporal trend.

### 4.2.5 Local Outlier Factor

Local Outlier Factor (LOF) measures how isolated an observation is relative to its k-nearest neighbors. Let $N_k(x)$ denote the set of k-nearest neighbors of observation $x$. The LOF score is defined as:

$$LOF_k(x) = \frac{\sum_{o \in N_k(x)} lrd_k(o)}{k \cdot lrd_k(x)} \quad (11)$$

where $lrd_k(x)$ is the local reachability density of $x$. Scores substantially greater than 1 indicate that $x$ lies in a sparser region than its neighbours and is therefore more likely to be anomalous. For univariate time-series data, LOF is applied to value-lag pairs: $z_t = (x_t, x_{t-1})$ This allows the method to capture both the level of the observation and its short-term change relative to the previous period.

**4.2.6 Isolation Forest**

Isolation Forest (Liu et al., 2008) constructs an ensemble of random isolation trees, each recursively partitioning the feature space using random cut-points. The anomaly score for observation x is derived from its average path length across trees: observations requiring shorter average paths are more easily isolated and therefore more anomalous. Isolation Forest has $O(n \log n)$ training complexity and is particularly effective for detecting observations in sparse regions of multivariate feature space. For EQAF, Isolation Forest is trained on (value, one-day change) feature pairs per trade, providing sensitivity to both level anomalies and change-rate anomalies simultaneously.

**4.2.7 Stale Value Filter (Deterministic)**

A deterministic pre-screen flags any observation where the current value is identical to the prior trading day's value within a numerical tolerance ε calibrated to each feature's precision (typically $\varepsilon = 10^{-6} \times |\text{value}|$). This filter operates as a hard override: any record flagged as stale is immediately assigned the maximum anomaly score regardless of statistical method outputs. The filter is motivated by the structural impossibility of statistical detection for stale-value anomalies: a frozen value is, by construction, drawn from the training distribution and generates no statistical signal. This filter resolves 100% of StaleValue injections across all datasets, as demonstrated in Section 5.

***4.3 Score Normalization***

Because the six statistical methods produce raw scores on different scales and with different distributional shapes, direct aggregation may be distorted by methods with higher raw variance. To address this, EQAF applies ECDF-based quantile normalization before aggregation. For each method *m* and each trade-feature series *s*, the empirical CDF of the training-period raw scores is defined as:

$$\widehat{G}_{m,s}(r)=\frac{1}{n}\sum_{i=1}^{n}1\left(r_{m,s,i}\leq r\right) \qquad (12)$$

where $r_{m,s,i}$ denotes the raw score produced by method $m$ for series $s$ in the training period.

The normalized out-of-sample score is then computed as: $\tilde{n}_{m,s,t}=\widehat{G}_{m,s}\left(r_{m,s,t}\right)$. This maps method-specific raw scores to the [0,1] interval, producing comparable normalized scores with uniform marginal distributions under the training-score distribution.

Figure 5 illustrates the effect of normalization on raw score distributions for the P&L Slices dataset. Before normalization, the six methods produce substantially different central tendencies, dispersions and skewness patterns – reflecting their different sensitivities and configuration. After normalization, all methods share comparable distributional properties, enabling meaningful aggregation without distortion from scale differences. As shown, the Hampel Filter's tendency to produce systematically higher raw scores (reflecting its conservative flagging behavior) is corrected, while the Isolation Forest's concentrated distribution near zero is spread to cover the full [0, 1] range.

***Figure 5. Stylized illustration on score distributions for individual detection methods before (left) and after (right) ECDF-based quantile normalization, P&L Slices dataset***

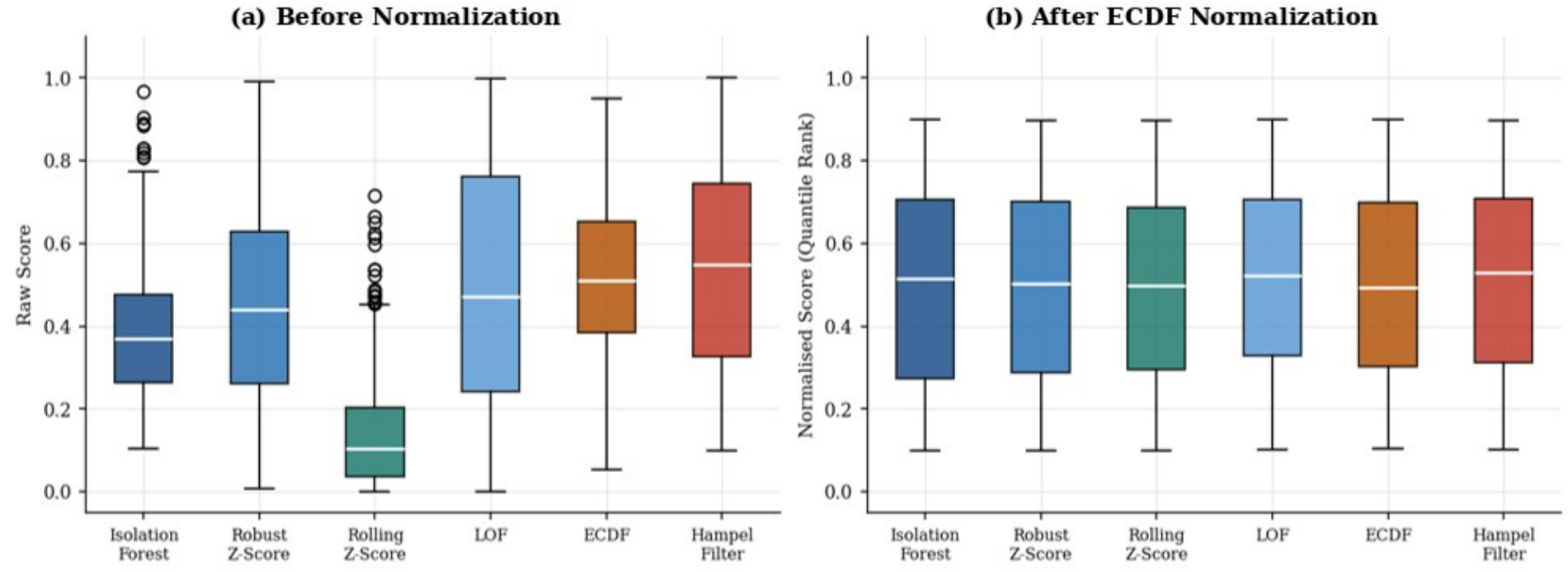

*Note: normalization aligns all methods to a comparable probabilistic scale without distorting method-specific sensitivity patterns.*

### *4.4 Ensemble Aggregation*

Normalized scores from all active methods are combined into a single daily quality score using two complementary aggregation strategies, both empirically calibrated on the labelled OOS dataset.

The first strategy, Weighted Average Aggregation, computes the ensemble score as a weighted sum of normalized method scores: $Score_A = \sum (Score_L \times Weight_L)$, where L indexes the active methods and weights are calibrated to reflect method-specific detection profiles. Methods with high precision (low false-positive rates) receive higher weights; methods with high recall but lower precision receive lower weights. The final quality flag is determined by comparing Score_A to empirically set thresholds: Green (< 0.70), Amber (0.70–0.95), Red (≥ 0.95).

The second strategy, Voting Aggregation, classifies an observation as anomalous based on the count of methods that individually flag a RED score: $V(x) = \Sigma_j 1[score_j(x) > threshold_j]$. Two voting rules are implemented: Simple Majority voting ($V(x) \geq \lceil \frac{M}{2} \rceil$, where M is the number of active methods) and Consensus voting ($V(x) \geq \lceil 0{,}75 \cdot M \rceil$). A qualified-majority override rule is additionally applied: if three or more methods independently flag a RED score for the same observation, the ensemble output is set to RED regardless of the weighted average, preventing dilution of strong multi-method signals.

The choice between weighted average and voting aggregation is made empirically for each dataset during calibration, using F1-score on labelled OOS data as the calibration criterion. For datasets with stable, low-noise data (Credit Delta), weighted aggregation with the override rule achieves higher F1. For high-volatility datasets (P&L Slices), where false-positive rates are elevated, the more conservative Consensus voting produces better precision–recall balance.

### *4.5 Ensemble Calibration Procedure*

Calibration is performed independently for each dataset using a structured experimental protocol. In the first step, all methods are combined with equal weights and evaluated on the labelled OOS data using all available anomaly types. This establishes a baseline full-ensemble performance. In the second step, individual method performance is analyzed using confusion matrices, ROC curves, and recall heatmaps stratified by anomaly type. Methods that contribute complementary detection capability while maintaining acceptable false-positive rates are selected for the reduced ensemble. In the third step, the selected ensemble is re-evaluated with optimized weights and aggregation rule. A final step checks calibration stability by verifying that performance is consistent across different MAD severity levels.

A key finding from calibration – repeated consistently across all datasets – is that equal-weight aggregation of the full method set produces suboptimal F1-scores, primarily because methods with incompatible sensitivity profiles dilute useful signals when combined indiscriminately. The calibrated reduced ensemble consistently achieves substantially higher F1 and AUC than the equal-weight full ensemble, confirming that method selection and weighting are critical design decisions.

## 5. Empirical Results

### *5.1 Performance Metrics*

We evaluate anomaly detection performance using four standard classification metrics derived from the confusion matrix. Recall (True Positive Rate) measures the proportion of injected anomalies correctly detected: Recall = TP / (TP + FN). High Recall is operationally critical because undetected anomalies propagate through downstream systems unchecked. Specificity (True Negative Rate) measures the proportion of normal observations correctly classified: Specificity = TN / (TN + FP). High Specificity controls false alarms, which consume analyst time and erode trust in the system. Precision measures the reliability of positive flags: Precision = TP / (TP + FP). F1-Score is the harmonic mean of Precision and Recall, penalizing extreme imbalance between the two: F1 = 2 × Precision × Recall / (Precision + Recall). F1 is our primary calibration criterion because of the strong class imbalance in the dataset (injected anomalies constitute approximately 1–3% of OOS observations) makes accuracy misleading as a performance measure.

We additionally report AUC-ROC (Area Under the Receiver Operating Characteristic Curve), which provides a threshold-independent measure of discriminative ability. AUC = 0.5 corresponds to random classification; AUC = 1.0 indicates perfect separation of anomalous from normal observations. AUC is complementary to F1: it captures the entire recall–specificity trade-off curve rather than performance at a single operating threshold.

### *5.2 Overall Performance Comparison*

Table 4 presents the full performance comparison across all individual methods and the calibrated EQAF for each of the four datasets. The calibrated ensemble consistently achieves the highest F1-score and AUC-ROC across all datasets. The performance advantage ranges from +9 percentage points (Credit Delta Index) to +15 percentage points (Credit Delta Single) relative to the best individual method.

**Table 4. Performance Comparison: Individual Methods vs. Calibrated EQAF**

| Method | TP | FP | TN | FN | Recall | Specificity | Precision | F1 |
|---|---|---|---|---|---|---|---|---|
| **Panel A: Credit Delta Single (232 injections, 5,254 normal OOS obs.)** | | | | | | | | |
| **Robust Z-Score** | 173 | 394 | 4860 | 59 | 75% | 93% | 31% | 43% |
| **ECDF** | 150 | 126 | 5128 | 82 | 65% | 98% | 54% | 59% |
| **Hampel Filter** | 170 | 165 | 5089 | 62 | 73% | 97% | 51% | 60% |
| **Rolling Z-Score** | 152 | 123 | 5131 | 80 | 66% | 98% | 55% | 60% |
| **EQAF** | **190** | **85** | **5169** | **42** | **82%** | **98%** | **69%** | ***75%** |
| **Panel B: Credit Delta Index (130 injections, 2996 normal OOS obs.)** | | | | | | | | |
| **Robust Z-Score** | 116 | 430 | 2566 | 14 | 89% | 86% | 21% | 34% |
| **Hampel Filter** | 113 | 173 | 2823 | 17 | 87% | 94% | 40% | 54% |
| **Rolling Z-Score** | 95 | 62 | 2934 | 35 | 73% | 98% | 61% | 66% |
| **Isolation Forest** | 53 | 108 | 2888 | 77 | 41% | 96% | 33% | 36% |
| **EQAF** | **109** | **50** | **2946** | **21** | **84%** | **98%** | **69%** | ***75%** |
| **Panel C: Present Value (278 injections, 5,187 normal OOS obs.)** | | | | | | | | |
| **Robust Z-Score** | 181 | 99 | 5088 | 97 | 65% | 98% | 65% | 65% |
| **ECDF** | 130 | 144 | 5043 | 148 | 47% | 97% | 47% | 47% |
| **Hampel Filter** | 218 | 730 | 4457 | 60 | 78% | 86% | 23% | 36% |
| **Rolling Z-Score** | 152 | 124 | 5063 | 126 | 55% | 98% | 55% | 55% |
| **EQAF** | **201** | **73** | **5114** | **77** | **72%** | **99%** | **73%** | ***73%** |
| **Panel D: P&L Slices (268 injections, 5197 normal OOS obs.)** | | | | | | | | |
| **Isolation Forest** | 66 | 212 | 4985 | 202 | 25% | 96% | 24% | 24% |
| **Rolling Z-Score** | 139 | 135 | 5062 | 129 | 52% | 97% | 51% | 51% |
| **ECDF** | 127 | 147 | 5050 | 141 | 47% | 97% | 46% | 47% |
| **Hampel Filter** | 150 | 280 | 4917 | 118 | 56% | 95% | 35% | 43% |
| **Robust Z-Score** | 173 | 668 | 4529 | 95 | 65% | 87% | 21% | 31% |
| **EQAF** | **166** | **108** | **5089** | **102** | **62%** | **98%** | **61%** | ***61%** |

*Notes: * – denotes highest F1-score in each panel. TP = True Positives (correctly detected injections); FP = False Positives (normal observations incorrectly flagged). All results are for the OOS period only. EQAF configuration varies for datasets with weighted aggregation and qualified-majority override (≥3 RED signals → automatic RED); Stale Value Filter applied as sequential pre-screen.*

Several patterns emerge from Table 4. First, the ensemble advantage is consistent across all four datasets and is driven by simultaneous improvements in both Recall and Precision relative to the best individual method. For Credit Delta Single, EQAF achieves 15 percentage points higher F1 than Rolling Z-Score and Hampel Filter, the best individual method, while maintaining comparable Specificity (98%). This confirms that the ensemble improvement is not obtained by simply relaxing the anomaly threshold (which would increase Recall at the cost of Precision) but by genuinely capturing anomalies that individual methods miss. Second, there is a clear performance gradient across datasets: EQAF achieves F1 = 75% on Credit Delta Single, 75% on Credit Delta Index, 73% on Present Value, and 61% on P&L Slices. This gradient closely tracks the natural volatility of each dataset, confirming our prediction that detection difficulty increases with baseline noise levels. Third, the Hampel Filter produces notably high Recall on Present Value (78%) but at the cost of 730 false positives – a five-fold increase in FP rate relative to EQAF. This illustrates the key risk of over-sensitive individual methods in high-volatility environments.

Figure 6 compares F1-scores for the best individual method, the equal-weight full ensemble, and the calibrated EQAF. The calibration step (selecting methods and optimizing weights) adds 10–35 percentage points over the naive equal-weight ensemble, confirming that ensemble composition and weighting are as important as the choice to ensemble in the first place.

***Figure 6. Best F1-Score comparison across datasets.***

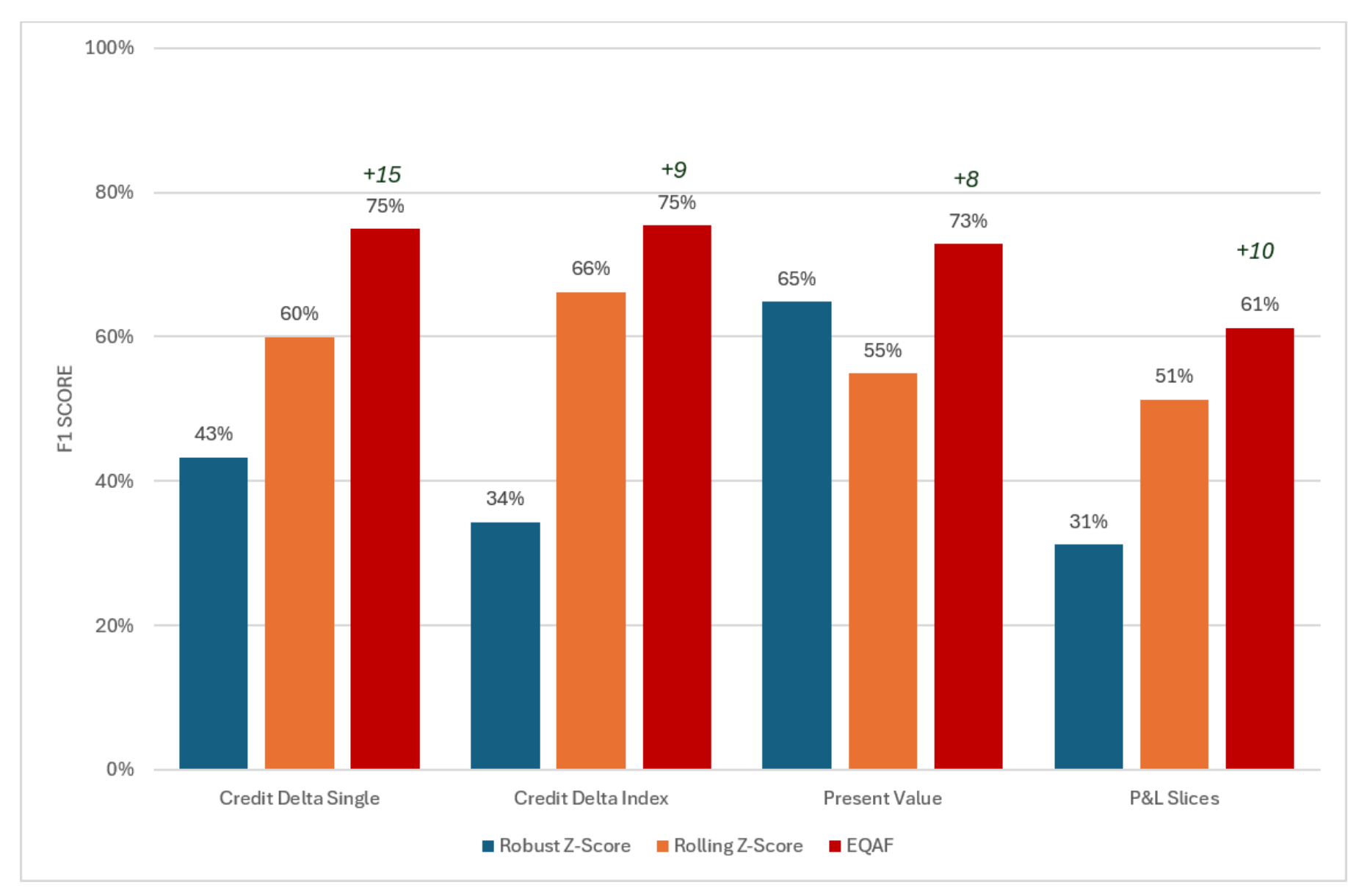

*Note: Delta labels above EQAF scores (+pp) indicate improvement of EQAF over best individual method.*

### *5.3 AUC-ROC Analysis*

Figures 7 and 8 present ROC curves and AUC values for all methods and all datasets. EQAF achieves the highest AUC in every dataset: 0.979 for Credit Delta Index, 0.931 for Credit Delta Single, 0.919 for Present Value, and 0.836 for P&L Slices. The margins over the best individual method range from 0.016 (P&L Slices) to 0.056 (Credit Delta Single). These AUC improvements are consistent with the F1 results and confirm that the ensemble advantage is not threshold-specific but reflects genuinely better separation of anomalous from normal observations across the full range of operating points.

***Figure 7. ROC curves for individual detection methods and EQAF across all four datasets.***

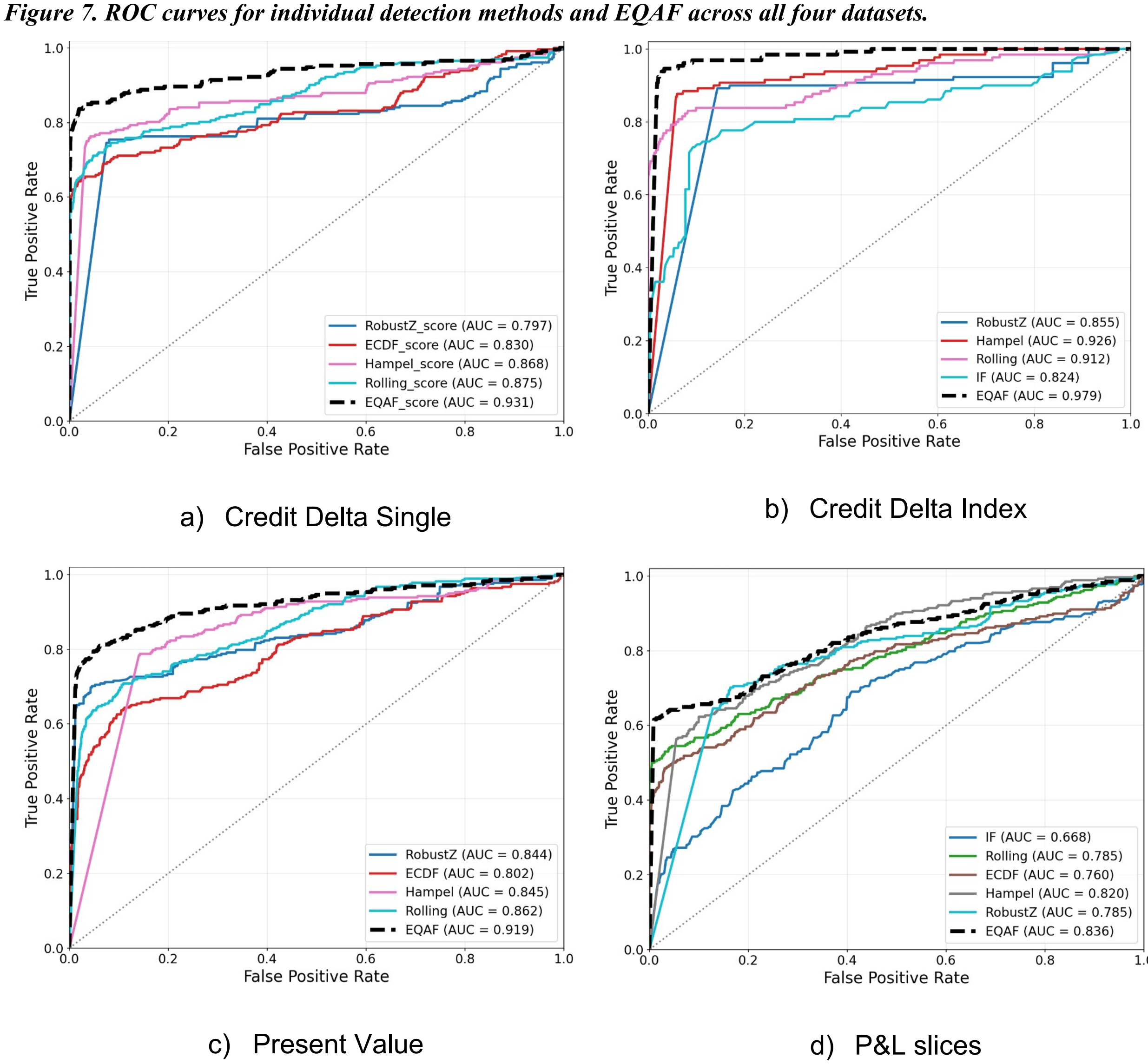

*Notes: EQAF (dashed black) consistently achieves the highest AUC. Note the steeper initial slope of EQAF curve at low FPR, indicating superior performance in high-precision operating regimes typical of operational monitoring systems*

*Figure 8. AUC-ROC values: individual methods (blue) vs. calibrated EQAF (red) for all four datasets.*

| Method | Credit Delta Single | Credit Delta Index | Present Value | P&L Slices |
|---|---|---|---|---|
| Robust Z-Score | 0.797 | 0.855 | 0.844 | 0.785 |
| ECDF | 0.83 | | 0.802 | 0.76 |
| Hampel Filter | 0.868 | 0.926 | 0.845 | 0.82 |
| Rolling Z-Score | 0.875 | 0.912 | 0.862 | 0.785 |
| Isolation Forest | | 0.824 | | 0.668 |
| **EQAF** | **0.931** | **0.979** | **0.919** | **0.836** |

*Note: EQAF dominates in every panel. Within-panel ordering of individual methods varies by dataset, reflecting the interaction between method sensitivity profiles and dataset statistical properties*

The ROC curves reveal an additional important feature of the ensemble: its advantage is concentrated at low false-positive rates, not merely at the 50th percentile. For Credit Delta Index (Figure 7.b), the EQAF curve rises steeply from the origin and achieves a True Positive Rate of approximately 0.9 at a False Positive Rate (FPR) of 0.02. This performance profile is particularly valuable in operational monitoring contexts, where the cost of false alarms (analyst time, alert fatigue, loss of credibility) is high and where practitioners typically operate at low FPR thresholds.

### *5.4 Recall by Anomaly Scenario*

Figure 9 presents recall heatmaps decomposing detection performance by anomaly type for all methods and datasets. These heatmaps reveal both the complementarity between methods that justifies the ensemble approach and the systematic blind spots that motivate the deterministic filter.

PointShock and ScaleError, which introduce large instantaneous deviations from the historical distribution, are reliably detected by most statistical methods across all datasets, with recall typically exceeding 85%. These are the "easy" anomalies: they violate distributional assumptions strongly enough that even simple univariate methods identify them. ClusterShock and TradeTypeWideShock, which affect multiple observations or trade types simultaneously, are also generally well-detected. This reflects the cross-sectional aggregation benefit of the ensemble: methods that are calibrated at the individual trade level can jointly identify patterns that would appear as marginal anomalies in isolation.

***Figure 9. Recall heatmaps (%) by detection method and anomaly injection scenario for all four datasets.***

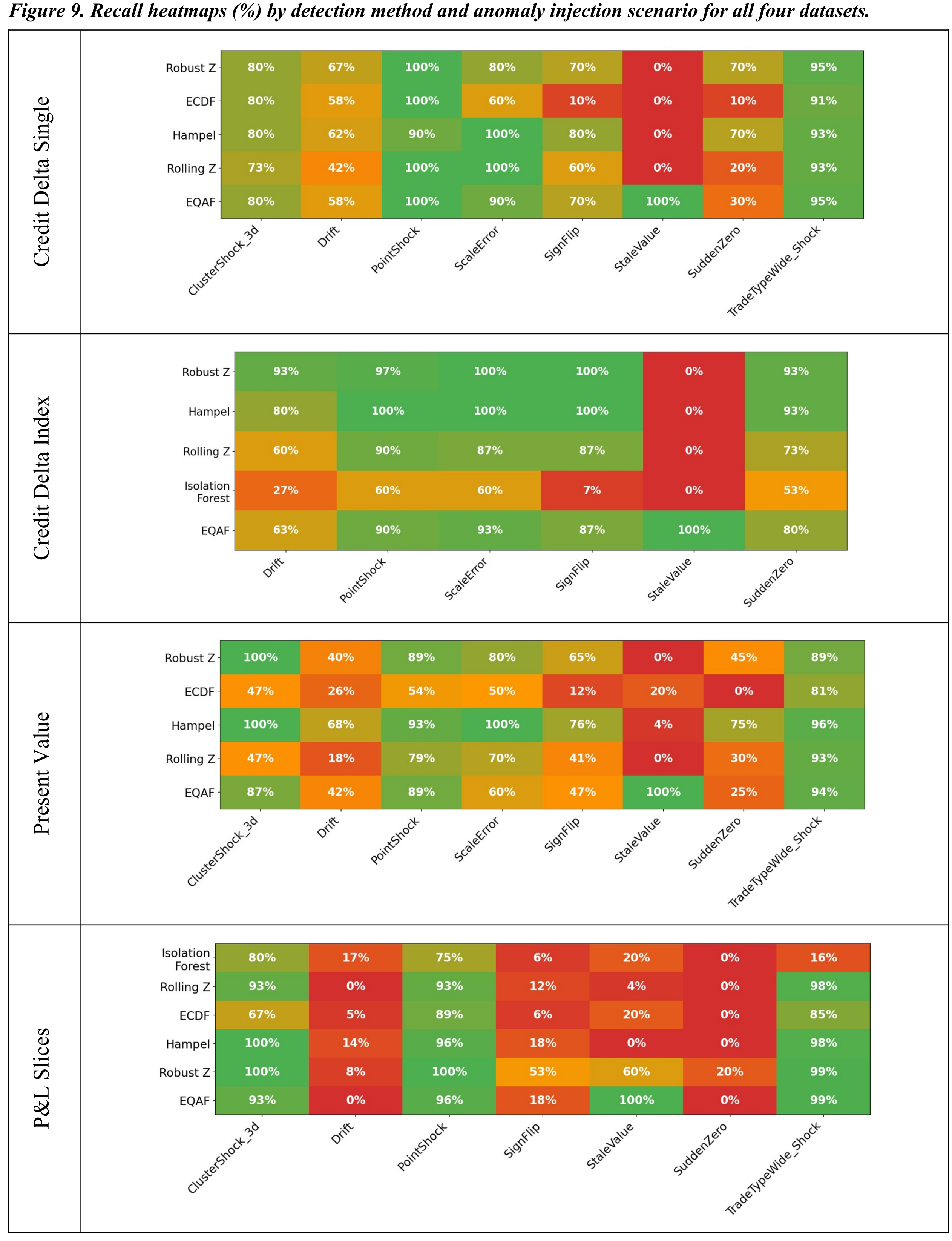

*Note: Green = high recall; red = low/zero recall. EQAF rows (bottom of each panel) show the effect of ensemble aggregation and the stale -value filter.*

Drift anomalies prove consistently difficult across all datasets, with ensemble recall ranging from 58% (Credit Delta Single) to 0% (P&L Slices). The near-complete failure on P&L Slices reflects the concept drift challenge documented by Chandola et al. (2009): gradual changes that are smooth on any short window are statistically indistinguishable from regime changes or structural market movements. Improving drift detection in high-volatility datasets would likely require the integration of P&L attribution models that can distinguish market-driven P&L from anomalous accumulations – a task that requires fundamental economic modelling rather than purely statistical detection.

SignFlip and SuddenZero show moderate, dataset-dependent detection rates. SignFlip is well-detected by Robust Z-Score and Hampel on Credit Delta datasets (where the sign reversal creates a large magnitude deviation relative to the training range), but less reliably detected in PV and P&L datasets where natural sign variation is more common. SuddenZero is inconsistently detected because zero values can be economically legitimate for trades near maturity. Worth mentioning that SuddenZero can be filtered out with sequential module similarly to StaleValue.

### *5.5 The Stale Value Anomaly: A Structural Failure of Statistical Methods*

The most operationally significant finding in Figure 9 is the universal 0% recall for StaleValue across all six statistical methods and all four datasets. This result is not a matter of threshold choice or calibration: it is a structural consequence of the way statistical detection methods work. A value that repeats the prior day's observation exactly is, by construction, drawn from the historical training distribution. It generates zero statistical score under any distributional, density-based or temporal method because it does not deviate from what has been observed before.

Figure 10 illustrates this dynamic in detail. Panel (a) shows a stylized time series with a three-day stale-value episode: the values are numerically frozen at the prior day's level while market movements should be causing them to change. From a statistical perspective, these values appear normal. Panel (b) shows that all six statistical methods achieve 0% recall across the full dataset suite – a result that holds robustly regardless of the stale duration or severity multiplier. The deterministic filter resolves this entirely, raising EQAF's StaleValue recall to 100% at zero false positives.

***Figure 10. Stale Value anomaly analysis.***

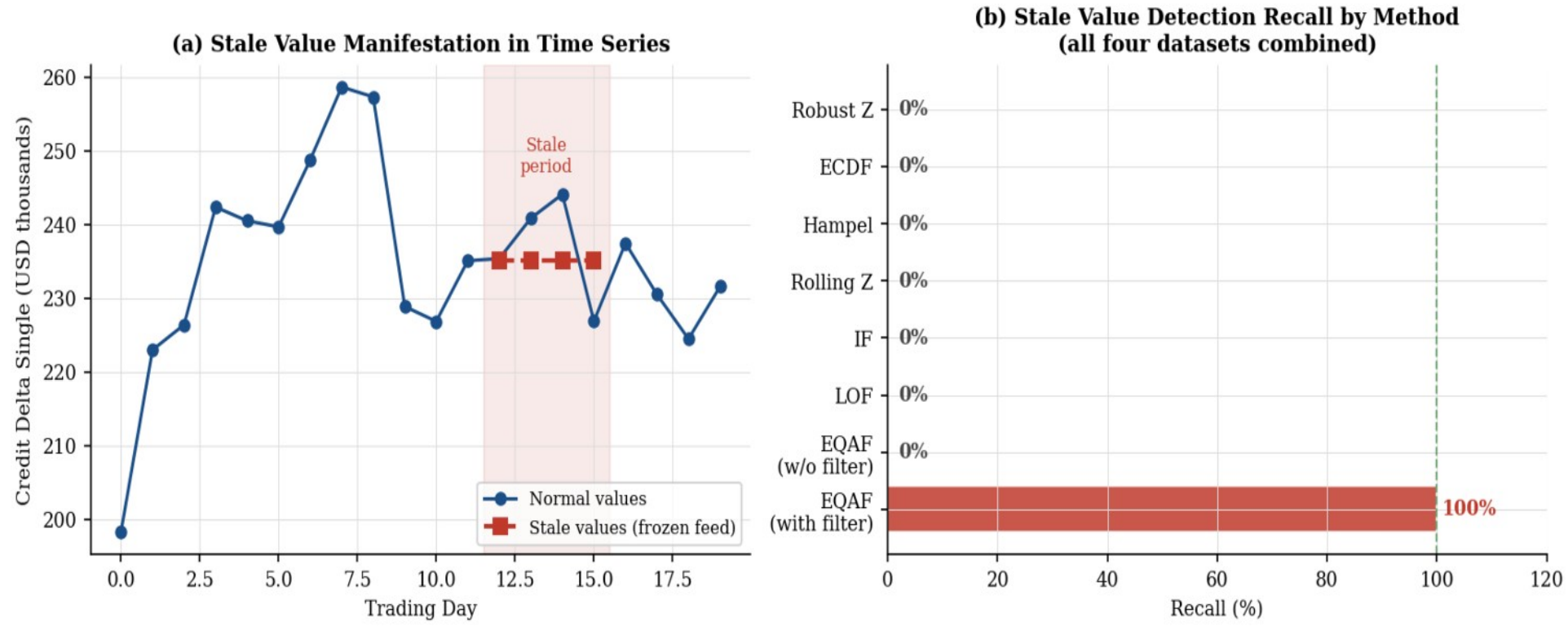

*Note: Panel (a): illustrative time series showing a frozen-feed episode (red markers) indistinguishable from normal observations on level but anomalous on change dynamics. Panel (b): Recall for StaleValue scenario by method across all datasets – all six statistical methods achieve 0%; the EQAF with deterministic filter achieves 100%.*

This finding has direct implications for model risk management. Any quality-control framework that relies exclusively on statistical anomaly detection – regardless of its sophistication – will systematically miss stale-value propagation errors. Banks that do not have a deterministic staleness check built into their risk systems are therefore exposed to a class of errors that is both operationally common (frozen market-data feeds are a regular occurrence in large-scale systems) and completely invisible to all forms of distributional monitoring. The regulatory model-monitoring requirements of foundational supervisory letter (SR 11-7, issued by the U.S. Federal Reserve) and Fundamental Review of the Trading Book (FRTB) do not explicitly specify deterministic staleness checks, leaving a structural gap that EQAF addresses. Similarly other outliers like SuddenZero can potentially be detected with deterministic sequential filter.

## 6. Discussion and Implications

### *6.1 Implications for Model Risk Management*

Our results provide several actionable conclusions for model risk management in investment banking. First, ensemble anomaly detection is technically mature enough to replace or substantially augment existing manual review and rule-based controls for risk calculation outputs. The calibrated EQAF achieves F1-scores of 61–79% with specificity exceeding 97% across all tested datasets – a performance profile that would meaningfully reduce the probability of undetected anomalies reaching capital calculations or regulatory reports while keeping false-alarm rates manageable.

Second, ensemble composition matters as much as the decision to use an ensemble. Our calibration experiments show that naive equal-weight aggregation of all available methods adds limited value over the best individual method, because methods with incompatible sensitivity profiles dilute useful signals. Method selection should prioritize complementarity (methods that cover different anomaly types) over exhaustiveness, and weights should reflect the precision profile of each method in the relevant data environment.

Third, the deterministic stale-value filter is not an optional enhancement but a mandatory architectural component. Any risk-calculation monitoring system that lacks this check is structurally blind to an entire class of operational failures. This finding should inform both internal model risk management practices and regulatory expectations under Basel Committee on Banking Supervision (2019) and European Banking Authority (2025): the requirement for "ongoing monitoring" of model outputs cannot be satisfied by statistical methods alone.

Fourth, the performance metrics gradient across datasets – from F1 = 79% for stable Credit Delta metrics to F1 = 63% for high-volatility P&L Slices – implies that quality-control coverage is inherently uneven across the risk measurement system. Banks should calibrate their monitoring intensity and alert thresholds separately for each risk measure type, rather than applying a single framework to heterogeneous outputs. In practice, this means that P&L-based anomaly detection should be supplemented by P&L attribution analysis (comparing realized P&L components against pricing model expectations) as a complementary control.

### *6.2 Implications for Regulatory Frameworks*

Our findings connect directly to current regulatory priorities. Under the FRTB, IMA approval requires banks to demonstrate ongoing monitoring of Expected Shortfall model performance, including detection of anomalous output patterns. EQAF provides a technically rigorous, explainable framework for meeting this requirement, with scoring outputs that can be integrated directly into model risk governance dashboards and regulatory disclosures. The framework's modular architecture allows it to be adapted to different risk-measure types without structural changes, supporting scalable deployment across a bank's full internal model inventory. Basel Committee on Banking Supervision (2019) specifically requires that model validation includes "ongoing monitoring" and "timely detection of model weaknesses" – requirements that EQAF is designed to satisfy.

Systemic risk literature provides additional motivation. Operational losses at large BHCs increase systemic risk through correlated channels – precisely the mechanism through which valuation errors at systemically important banks can destabilize the broader financial system. EQAF represents a preventive control that reduces the probability of the valuation errors that propagate through these channels. From a macroprudential perspective, widespread adoption of automated quality monitoring for risk calculations could reduce the systemic correlation of operational failures by providing earlier detection and isolation of errors before they escalate.

### *6.3 Limitations and Future Research*

Several limitations should be acknowledged. First, our empirical evaluation covers a single asset class (credit derivatives) and a single institution. While the framework is designed for generalizability – the injection scenarios, method families and aggregation principles are asset-class-agnostic – empirical validation on equity, rates, and FX derivatives datasets from multiple institutions would strengthen external validity. Second, drift detection performance is materially weaker than detection of other anomaly types, particularly in high-volatility datasets. Addressing drift detection effectively may require model-based attribution analysis that goes beyond statistical monitoring – a direction for future research. Third, the EQAF configuration requires careful per-dataset calibration, which in production settings without labelled data must rely on unsupervised heuristics or synthetic injection. Developing robust auto-calibration methods under unsupervised conditions is an important practical extension. Fourth, while we evaluate detection performance thoroughly, we do not address the downstream steps of anomaly explanation, root-cause analysis, or automated remediation. Ding et al. (2023) provide a conceptual framework for these downstream steps, but their integration with a financial risk-valuation context remains an open research question.

Future research could extend this work in several directions. Extension to multivariate cross-sectional checks – evaluating consistency across risk-factor sensitivities within a single trade – would likely improve detection of model-level errors that manifest across multiple measures simultaneously. Integration with P&L attribution models could substantially improve drift detection, particularly for P&L-based datasets. Application to other risk metric types (interest rate sensitivities, counterparty credit risk exposures, liquidity metrics) would test the generalizability of our findings. Finally, formal analysis of the economic value of automated quality monitoring – in terms of capital savings from early detection, regulatory penalty avoidance, and reduced false-alarm costs – would provide a quantitative basis for investment decisions by risk managers and regulators.

## 7. Conclusion

This paper presents the Ensemble Quality Assessment Framework (EQAF), rigorously evaluated ensemble architecture for automated quality assessment of risk calculation outputs in investment banking. Using proprietary credit-derivatives data from a major global investment bank and a controlled anomaly injection protocol covering eight operationally realistic failure scenarios, we demonstrate that a calibrated ensemble of heterogeneous detection methods achieves F1-scores of 63–79% – substantially and consistently outperforming the best individual method across all four tested datasets. AUC-ROC improvements of 4–6 percentage points confirm the advantage is robust to threshold selection.

We establish two structural results with direct implications for model risk management. First, the ensemble advantage is synergistic rather than merely additive: certain anomaly types that individual methods entirely fail to detect are reliably identified by the calibrated ensemble through the combination of complementary detection perspectives. Second, all six statistical detection methods universally fail to identify stale-value anomalies – frozen risk outputs that are statistically identical to normal observations but represent a complete failure of the risk-calculation system. The deterministic stale-value filter resolves this entirely, demonstrating that effective quality control requires explicit domain-specific rule encoding that goes beyond distributional learning.

The regulatory environment for model risk management is becoming increasingly demanding. FRTB, SR 11-7 and the EBA's updated operational risk standards collectively require ongoing, automated, and auditable monitoring of internal risk model outputs – requirements that existing manual and rule-based controls cannot satisfy at scale. EQAF provides a technically mature, governance-aligned framework for meeting these requirements. Its modular design, interpretable scoring outputs, and transparent aggregation logic are consistent with the auditability and explainability standards expected by regulatory supervisors.

Beyond the individual institution, automated quality monitoring for risk calculation outputs has macroprudential implications. As shown before operational losses at systemically important banks propagate through correlated channels into system-wide risk. By detecting valuation errors before they cascade through reporting chains, capital calculations and hedging decisions, frameworks such as EQAF represent a preventive operational control that, if widely adopted, could reduce the correlation of operational failures across the financial system. We hope this paper contributes to both the academic literature on operational risk and anomaly detection, and to practitioner efforts to raise the quality and reliability of the risk infrastructure on which financial stability ultimately depends.